\documentclass{article}

\usepackage{graphicx}
\usepackage{fullpage,amsmath,amssymb,latexsym}
\usepackage{multirow}
\usepackage{natbib}

\begin{document}

\title{The Change in Jupiter's Moment of Inertia due to Core Erosion and Planetary Contraction}

\author{Ravit Helled\\
Department of Geophysics and Planetary Sciences\\
   Tel-Aviv University, Israel
}

\date{}
\maketitle 

\begin{abstract}
We explore the change in Jupiter's normalized axial moment of inertia (NMOI) assuming that Jupiter undergoes core erosion. It is found that Jupiter's contraction combined with an erosion of 20 M$_{\oplus}$ from a primordial core of 30 M$_{\oplus}$ can change Jupiter's NMOI over time significantly. It is shown that Jupiter's NMOI could have changed from $\sim$ 0.235 to $\sim$ 0.264 throughout its evolution. 
We find that a NMOI value of $\sim$0.235 as suggested by dynamical models (Ward \& Canup, 2006, ApJ, 640, L91) could, in principle, be consistent with Jupiter's primordial internal structure. Low NMOI values, however, persist only for the first $\sim 10^6$ years of Jupiter's evolution. Re-evaluation of dynamical stability models as well as more sophisticated evolution models of Jupiter with core erosion seem to be required in order to provide more robust estimates for Jupiter's primordial NMOI. 
\end{abstract}


\section{Introduction}

A determination of Jupiter's core mass has important implications for our understanding of giant planet origin. Since the existence of a core of $\sim 10$ Earth masses (M$_{\oplus}$) in Jupiter would provide an observational support for the standard theory for giant planet formation, the {\it core accretion} model (Pollack et al., 1996), a great effort to better constrain Jupiter's core mass has been carried. Space missions to Jupiter have been sent to determine its gravitational field, with {\it Juno} currently being on its way to Jupiter with the goal of providing accurate measurements of Jupiter's gravitational harmonics up to $\sim$J$_{12}$ (Bolton, 2006). Detailed interior models of Jupiter have been developed to better constrain its internal structure, with a clear focus on its core masses (Nettelmann et al., 2008, Militzer et al., 2008), and alternative methods to determine the core mass of Jupiter have been investigated and developed (Gaulme et al., 2011). \par

It  is possible, however, that Jupiter's core mass today is different from its primordial one; Jupiter's core mass shortly after formation could have been significantly larger than Jupiter's current core mass due to core erosion (Stevenson, 1982; Guillot et al., 2004; Wilson and Militzer, 2011a,b). From recent measurements of Jupiter's gravitational field, Jupiter's core mass is estimated to be between zero and ten M$_{\oplus}$ (Saumon and Guillot, 2004). 
If core erosion indeed occurred in Jupiter, the core mass that is determined from interpretation of gravity data of Jupiter today should {\it not} be directly related to Jupiter's origin without a consideration of the evolution of Jupiter's internal structure. 
If Jupiter's core eroded over time, with the primordial core being significantly more massive (e.g., 30 M$_{\oplus}$), it would undoubtedly support the {\it core accretion} mechanism. In addition, a massive primordial core will remove the difficulty of forming Jupiter's within a few million years. This is due to the fact that a more massive core results in runaway gas accretion at an earlier time, and therefore decreasing the total formation timescale (Pollack et al., 1996; Hubickyj et al., 2005). Furthermore, with a massive primordial core, the accreted gas can be depleted with heavy elements and still lead to the formation of Jupiter with heavy-element enrichment as predicted by interior models, since the heavy elements from the core would mix with the gaseous envelope.  \par 

The possibility of  core erosion has implications to our understanding of gas giant planets in general.  Estimates of giant planets' core masses are often used to discriminate between giant planet formation models, and to put constraints on the physical properties of the protoplanetary disks in which the planets are formed. However, since core erosion can occur, estimates of core masses of giant planets should be taken with caution, as the derived core masses may not be equal to the primordial core masses.
\par

The question of whether Jupiter's core has been eroded is still open, and due to the complexity of the subject detailed investigations are required before this question can be answered (e.g., Wilson and Militzer, 2011a). 
In this note, we do {\it not} model (or investigate the feasibility of) core erosion in Jupiter. Instead, we investigate how core erosion, if indeed occurred in Jupiter, could affect its internal structure over time, in particular, its normalized moment of inertia (NMOI$\equiv C/MR^2$, where $C$ is the axial moment of inertia and $M$ and $R$ are the planetary total mass and mean radius, respectively). The motivation of this work is to explore whether Jupiter's NMOI could have been significantly lower in its early evolution to fit  the low NMOI value of $\sim$0.236 as proposed by Ward \& Canup (2006, hereafter WC06) based on dynamical considerations that link between Jupiter's obliquity and the precession of Uranus' orbit plane. Helled et al. (2011) have investigated the possible range of NMOI values for Jupiter; it was found that Jupiter's NMOI is between 0.2629 and 0.2645. Although a large range of interior models was investigated, no solutions that reproduce the low NMOI value derived by WC06 could be found. The models of Helled et al., (2011), however, describe Jupiter's internal structure at present. 
Below, using a simple model, we derive Jupiter's NMOI as a function of time when core erosion and planetary contraction are included.

\section{Jupiter's NMOI vs. time using a simple model}

Guillot et al.\ (2004) have presented an approximate approach to quantify the erosion of Jupiter's core based on the gravitational potential energy cost of lifting the core material into the gaseous envelope. 
In this work we adopt the core erosion rate $dM_C/dt$ where $M_C$ is Jupiter's core mass,  of Guillot et al.\ (2004), as well as their initial and final core masses of 30 M$_{\oplus}$ and 11 M$_{\oplus}$, respectively. 
Due to the simplicity of the  core erosion rate given by Guillot et al. (2004), and in order to investigate the sensitivity of the change in NMOI over time to the assumed core erosion rate, we also consider other core erosion rates. 
\par
 
While Jupiter's total mass has not changed throughout its evolution, its radius has been decreasing with time. As a result, we also account for the change in Jupiter's radius with time $dR/dt$ when deriving the NMOI vs. time. 
It is important to include the planetary contraction due to its strong effect on the planetary NMOI. At early times, the planet is young with a thin and extended gaseous envelope (ideal gas), with most of the mass being concentrated toward the planetary center. Under these conditions, the NMOI is fairly low. As the planet contracts and evolves, the material inside the planet becomes more degenerate, and the planet becomes less centrally condensed. While the dimensional moment of inertia value decreases with time due to the planetary contraction, it does it with a slower rate than $MR^2$, and as a result, the net effect is an increase in the planetary NMOI. In order to isolate the effect of the planetary contraction on Jupiter's NMOI with time, we also consider two cases in which the core masses stay unchanged as Jupiter evolves.\par


We consider six different cases for the core erosion rate, always accounting for the planetary contraction. 
For consistency, the four cases in which the core mass is changing with time, we start with a primordial core mass of 30 M$_{\oplus}$ and a core mass of 11 M$_{\oplus}$ at present (Guillot et al., 2004). 
The cases we consider are as follows: (1) core erosion rate of Guillot et al. (2004) (2) core erosion rate linear in Log(t) (3) core erosion rate quadratic in Log(t) (4) core erosion rate proportional to Log(t)$^{(1/2)}$ (5) constant core mass of 30 M$_{\oplus}$, and (6) constant core mass of 10 M$_{\oplus}$. Figure 1 shows the different core erosion rates we consider. The small panel in the upper right shows the change in Jupiter's radius over time that we adopt for this work taken from Burrows et al. (1997). 
\par

Given that the core erosion rate is essentially unknown and the fact that there are no observational constraints for Jupiter's internal structure in its early state and as it evolves, we use a simple model to describe Jupiter's interior in which the density profile is described by a constant density core and a linear-density envelope (i.e., of the form $\rho(r)=\alpha(1-r)$ where $r$ is Jupiter's normalized radius). The advantages in using this interior model is that Jupiter's internal structure consists of only one free parameter ($\alpha$) for a given core mass and radius, and that despite its simplicity, a linear density profile for Jupiter can still reproduce Jupiter's gravitational coefficients fairly well (Zharkov and Trubitsyn, 1978).

We first search for an interior model for Jupiter at present with a core mass of 11 M$_{\oplus}$ and search for the core density and envelope density that best reproduce Jupiter's measured J$_2$. Given a core mass and radius as a function of time, the coefficient that defines the envelope density can then be found using the constraint that the total planetary mass must be equal to Jupiter's mass. For Jupiter at present we find a core density of 5.3 g cm$^{-3}$ and $\alpha\sim 4$. With these parameters Jupiter's J$_2$ is found to be 0.014699 with NMOI of 0.264705. In comparison, Jupiter's measured J$_2$ is 0.0146966 with a NMOI of 0.264705 using the Radau-Darwin approximation. Note that these values are normalized to Jupiter's mean radius today which is set to be 69,893.175 km (Helled et al., 2011). \par

The change in core mass over time $dM_C/dt$ determines $M_C(t)$. We then assume that Jupiter's core density has not changed as Jupiter's evolved, while the envelope's density has been changing with time as the core is eroding and contributing heavy-element material to the gaseous envelope; we then derive Jupiter's core radius over time $R_C(t)$. For a given time, core mass, core radius, planetary radius, and Jupiter's total mass ($M$=1.89815 $\times 10^{30}$g), we compute the corresponding envelope density profile vs. time and derive the corresponding NMOI for the six different cases. 

The theory of figures (Zharkov \& Trubitsyn, 1978) is used to derive Jupiter's interior structure, in which Jupiter's NMOI and J$_2$ are derived to third order in the small rotational parameter. Jupiter's rotation rate is held constant to its current value of 9h 55m 29.71s while Jupiter's mean radius is scaled using the contraction rate of Burrows et al. (1997). For a given time and core mass, we compute the corresponding envelope density profile and NMOI as a function of time. 
The results for Jupiter's calculated NMOI are shown in figure 2. \par

From looking at the cases with constant core masses, it is clear from the figure that indeed the planetary contraction alone results in an increase of the NMOI over time. For a core mass of 30 M$_{\oplus}$ it is found that Jupiter could have had low NMOI values during its early evolution, however, the derived NMOI for today (0.259) is inconsistent with Jupiter's NMOI of 0.263-0.2645 (Helled et al., 2011). A constant core mass of 10 M$_{\oplus}$ can reproduce NMOI factor that is consistent with Jupiter's gravity field, on the other hand, but cannot reach NMOI of 0.23-0.24 even at young ages of $10^5-10^6$ years due to the low core mass. The minimum NMOI value for this case is 0.246. Clearly, lower core masses will result in even higher NMOI values, while more massive cores will lead to lower NMOI values.\par   

For the cases with core erosion, we start with a fairly massive core (30 M$_{\oplus}$) and a final core mass of 11 M$_{\oplus}$ at present (Guillot et al., 2004). 
For these cases we get a final NMOI value that is consistent with Jupiter's gravity data, and in addition, primordial NMOI values that are consistent with low NMOI values (WC06).
When erosion of 20 M$_{\oplus}$ from Jupiter's core is included, Jupiter's NMOI value changes by more than 12\% during its evolution. The difference in the derived NMOI values vs. time for the different core erosion rates we consider is fairly small. The NMOI of Jupiter at present is found to be $\sim$ 0.2637, a value which is in a very good agreement with the results of Helled et al. (2011), while Jupiter's NMOI during its early evolution is found to be $\sim$ 0.2351 significantly smaller. During the first $\sim 10^6$ years of Jupiter's evolution, with a massive core and an extended envelope, the NMOI can be as low as $\sim$0.235, similar to the value suggested by WC06. However, this timescale is  fairly short compared to the time required to affect Jupiter's obliquity by perturbations from Uranus (WC06). \par

\subsection{Planetary Contraction}
When Jupiter's contraction is not included the change of the NMOI during Jupiter's evolution is significantly smaller and ranges between 0.2614 and 0.2647, and cannot lead to a large range of NMOI values. The {\it combination} of an extended envelope with a massive core is therefore required to obtain low NMOI values for Jupiter at young ages. 

Jupiter's radius at early times, however, is not well constrained, and this uncertainty results in an uncertainty in Jupiter's primordial NMOI. Larger primordial radii will lead to even lower NMOI values, while more compact configurations will increase the NMOI. The largest uncertainty in Jupiter's radius, however, persists for only $\sim10^7$ years (Marley et al., 2007). A better understanding of giant planet formation and early evolution could put useful constraints on the planetary radius at early stages, and therefore, on Jupiter's primordial NMOI. \par

It should be noted that the process of core erosion itself can affect the planetary contraction rate, and therefore in order to better constrain Jupiter's NMOI in its early state self-consistent evolution models must be developed. 
Planets with more massive cores tend to have smaller radius (Baraffe et al., 2008) and as a result, core erosion could lead to an increase in Jupiter's radius, and a decrease in NMOI. The concentration and distribution of heavy elements in the envelope also affect the NMOI and it is therefore important to follow the heavy elements as they leave the core and get distributed in the gaseous envelope.

\subsection{The Density of the Core}
The models presented above used core densities of 5.3 g cm$^{-3}$ since this density was found to best reproduce Jupiter's current J$_2$. However, this core density is low compared to core densities that are derived by interior models. Typical core densities for Jupiter are found to be around 10 g cm$^{-3}$ and could even reach 20 g cm$^{-3}$ (Guillot, 2005; Nettelmann et al., 2008). Since a core density of 5.3 g cm$^{-3}$ is low compared to densities from physical interior models we repeat the calculation assuming two other core densities: 10 and 20 g cm$^{-3}$.
The results are presented in figure 2 in panels (b) and (c), respectively. The derived NMOI at present is found to be about 0.262 and 0.260 for core densities of 10 and 20 g cm$^{-3}$, respectively. The timescale in which the NMOI is $\sim$0.235 remains of the order of $10^6$ years, and we can therefore conclude that within the limits of the model assumptions our results are robust.

\subsection{A Power-law Density Profile for the Envelope}
Despite the good fit of the linear-envelope model to Jupiter's measured J$_2$, a linear envelope's density is inconsistent with the behavior of hydrogen and helium, and a power-law for the envelope density is somewhat more realistic (Podolak and Hubbard, 1998). We therefore repeat the calculation using a density profile of the form 
$\rho(r)=\alpha (1 - r)^{\gamma}$, and following Podolak and Hubbard (1998) set $\gamma=0.7$ for Jupiter at present. Again, we find the core density to best reproduce Jupiter's measured J$_2$ and vary the density profile as the core mass is decreasing and the planet is contracting. The fit to Jupiter's measured J$_2$ value with a core mass of $\sim$11 M$_{\oplus}$ is not as good as in the linear case, and in addition, the power-law model is less constrained due to the existence of two free parameters for the envelope density profile. Nevertheless, the results are similar to the ones found for the linear-density case.  Again, it is found that at early times Jupiter's NMOI could have been significantly lower than its present-day value.

\subsection{Other Core Erosion Rates}
All the core erosion rates we consider are roughly logarithmic in time. Other core erosion rates, however,  are also possible, and we therefore also consider a linear-in-time core erosion rate in order to investigate the sensitivity of the results to the assumed erosion rate.  
For a linear-in-time erosion rate the decrease in the core mass that occurs at early times is smaller, and as a result, the timescale for low NMOI value is slightly lengthen. However, the difference from the cases in which the erosion rate is logarithmic in time is very small and the results are essentially the same. Similarly to the other four cases in which core erosion is included, Jupiter's NMOI is found to be low at first, and to increase to a value of $\sim$ 0.264 at present. Jupiter's NMOI is found to be about 0.242 and 0.248 at 10$^6$ and 10$^7$ years, respectively, for the linear-in-time erosion rate. 
Any core erosion rates with primordial core mass of 30 M$_{\oplus}$ and final core mass of 10 M$_{\oplus}$ will fall between the two curves in figure 2 that represent the the constant core masses.

\subsection{Primordial and Current Core Masses}
All the core erosion rates we consider in this work are somewhat similar. Although they vary differently with time, they all have the same initial and final masses for Jupiter's core. The two cases with constant core masses suggest that a primordial core mass of about 30 M$_{\oplus}$ is required to allow NMOI values of the order of 0.235 while small cores are required to fit NMOI values that are consistent with Jupiter's gravity data. There are two major points that should be kept in mind: (1) These core masses correspond to the simple interior model of a constant-density core and a linear envelope we used for the analysis. More complex internal structure could lead to different values since the NMOI depends on the overall distribution of material within the planet. (2) There is no requirement that the low core mass (10 M$_{\oplus}$) is reached at timescales of the order of $10^9$ years. If the initial model remains with a massive core but 20 Earth masses are eroded from the core  within shorter timescales (e.g., 10$^6-10^8$ years) the results will be essentially the same. The key conclusion of this work is that we can get a substantial change in Jupiter's NMOI value when {\it both} significant core erosion and planetary contraction are included. There are no requirements or constraints on {\it when} or {\it how long} the process of core erosion must occur. Clearly, other values for both the primordial and current core masses are possible, and we hope that further investigations of the topic would lead to more accurate determination of these values.

\section{Summary and Conclusions}
We investigate the possible change in Jupiter's NMOI over time due to core erosion and planetary contraction, using a core+envelope model for Jupiter's interior. We assume a primordial core mass of 30 M$_{\oplus}$ and an erosion of 20 M$_{\oplus}$ throughout Jupiter's evolution.
We consider three different core densities and various core erosion rates. 
We find that Jupiter's NMOI value could have changed by more than 12\% throughout its evolution, and that in its early state, due to the existence of a massive core (30 M$_{\oplus}$) and an extended envelope its NMOI could have been significantly lower, similar to the value predicted by WC06, but only for relatively short timescales.  This result corresponds to all the core densities and core erosion rates we consider, and we suggest that further investigation of the topic is required before this inconsistency can be resolved.
While it is still not possible to determine whether core erosion indeed occurred in Jupiter, recent investigations of the topic suggest that it is certainty feasible. Recent studies by Wilson and Militzer (2011a, b) suggest that a significant mass of Jupiter's core could have been eroded. \par

Despite its simplicity, our work demonstrates the importance of accounting for the change in Jupiter's interior structure as it evolves, and emphasizes the caution that has to be taken when relating Jupiter's current structure and its origin.  Measurements of Jupiter's gravitational field correspond solely to Jupiter at present, and we must study the possible evolutionary scenarios that could lead to its current state, in particular, its current core mass.  
The data provided by {\it Juno} may improve our understanding of Jupiter's origin, evolution, and internal structure. However, in order to take a full advantage of the data, and to interpret the measurements correctly, detailed investigations of core erosion and remixing processes must be carried.  We hope that our work will encourage further development of evolution models of Jupiter and a cautious approach when relating Jupiter's current interior structure and its origin. This caution should also be taken when estimated core masses of giant extrasolar planets are applied to discriminate among giant planet formation models.

\section*{acknowledgments}
I thank an anonymous referee for fruitful and valuable comments. 

\section*{References}
Baraffe, I., Chabrier, G., Barman, T., 2008. A\&A, 482, 315.\\
Bolton, S., 2006. European Planetary Science Congress 2006.,535.\\
Burrows, A., Marley, M., Hubbard, W. B., Lunine, J. I., Guillot, T., Saumon, D., Freedman, R., Sudarsky, D., Sharp, C., 1997. ApJ, 491, 856. \\
Helled, R., Anderson, J. D., Schubert, G. and Stevenson, D. J., 201. Icarus, 216, 440.\\
Hubickyj, O., Bodenheimer, P., and Lissauer, J. J., 2005. Icarus, 179, 415.\\
Gaulme, P., Schmider, F. X., Gay, J., Guillot, T. and Jacob, C. 2011. Astronomy \& Astrophysics, Volume 531, id.A104\\
Guillot, T., 2005. Annu. Rev. Earth Planet. Sci., 33, 493.\\
Guillot, T., Stevenson, D. J., Hubbard, W. B. and Saumon, D., 2004.  In Jupiter: The Planet, Satel lites and Magnetosphere, eds. F. Bagenal, T. E. Dowling and W. B. 
McKinnon, Cambridge Planetary Science, Cambridge, UK: Cambridge University Press,  35.\\
Marley, M. S., Fortney, J. J.; Hubickyj, O., Bodenheimer, P. Lissauer, J. J., 2007. ApJ, 655, 541.\\
Militzer, B., Hubbard, W. B., Vorberger, J., Tamblyn, I. and Bonev, S. A., 2008. The Astrophysical Journal, 688, L45.
Nettelmann, N., Holst, B., Kietzmann, A., French, M., and Redmer, R., 2008. The Astrophysical Journal, 683, 1217. \\
Podolak, M. and Hubbard, W. B., 1998. Solar System Ices. Dordrecht Kluwer Academic Publishers, Astrophysics and space science library (ASSL) Series vol no 227. ISBN0792349024., 735.\\
Pollack, J. B., Hubickyj, O., Bodenheimer, P., Lissauer, J. J., Podolak, M., and Greenzweig, Y., 1996. Icarus, 124, 62.\\
Saumon, D. and Guillot, T., 2004. The Astrophysical Journal, 609, 1170.\\
Stevenson, D. J., 1982. Planet. Space. Sci. 30, 755. \\
Ward, W. R. and Canup, R. M., 2006. The Astrophysical Journal, 640,  L91.\\
Wilson, H. F. and Militzer, B., 2011a. eprint arXiv:1012.5413.\\
Wilson, H. F. and Militzer, B., 2011b. eprint arXiv:1111.6309.\\
Zharkov, V. N. and Trubitsyn, V. P. 1978, Physics of planetary interiors (Astronomy and Astrophysics Series, Tucson: Pachart, 1978).\\

\clearpage

\begin{figure}
\begin{center}
 \includegraphics[width=7.in]{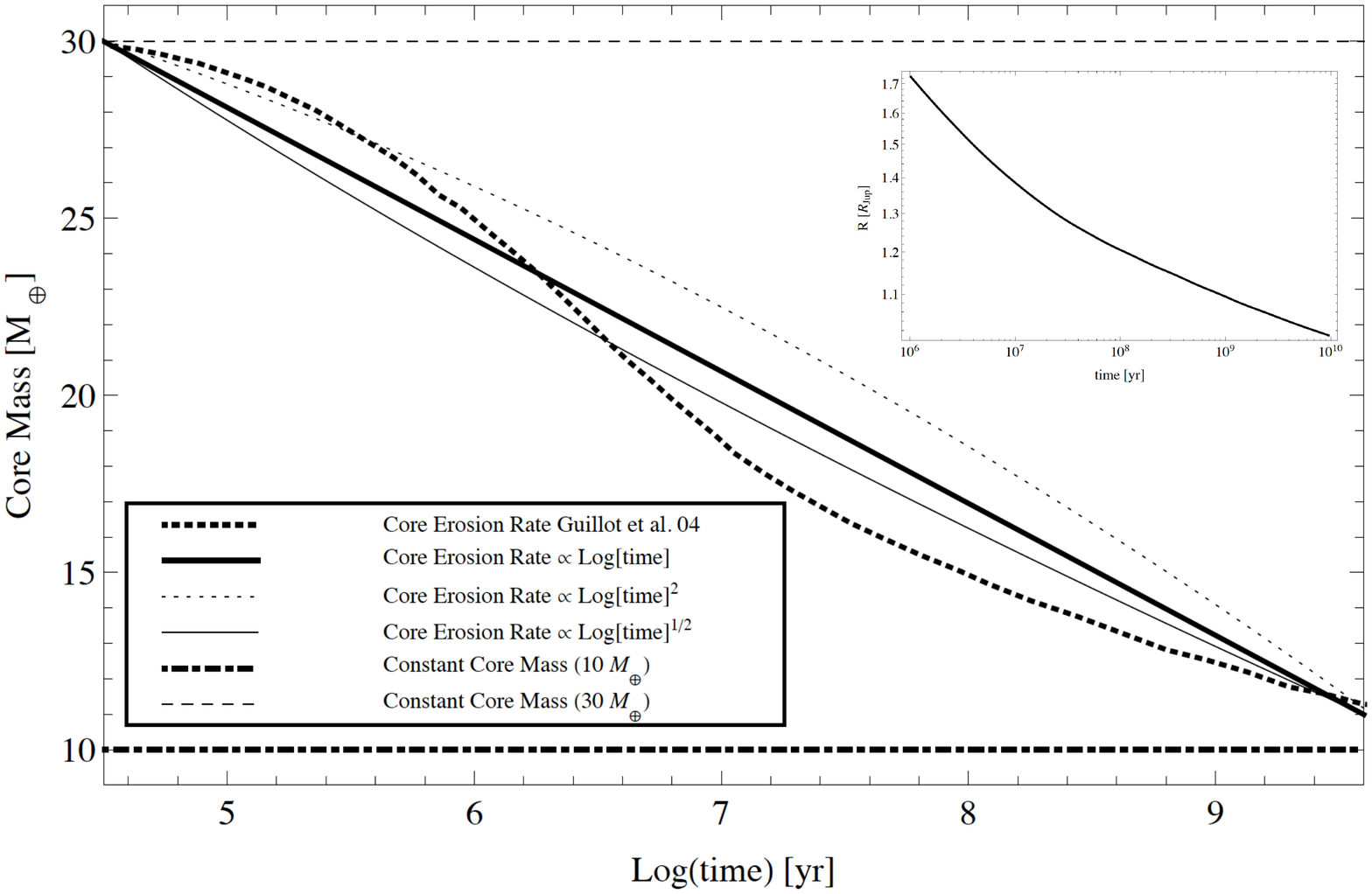} 
\caption{Jupiter's core mass as a function time. We consider six different erosion rates: (1) Guillot et al. (2004) core erosion rate: thick dotted curve, (2) core erosion rate linear in Log[time]: thick solid line, (3) core erosion rate quadratic in Log[time]: thin dotted line, (4) core erosion rate proportional to Log[time]$^{1/2}$: thin solid line, (5) constant core mass of 10 M$_\oplus$: thick dashed-dotted line, and (6) constant core mass of 30 M$_\oplus$: thin dotted line. The small panel on the top right shows Jupiter's radius vs. time from Burrows et al., (1997). }
\end{center}
\end{figure}

\clearpage

\begin{figure}
\begin{center}
 \includegraphics[width=4.15in]{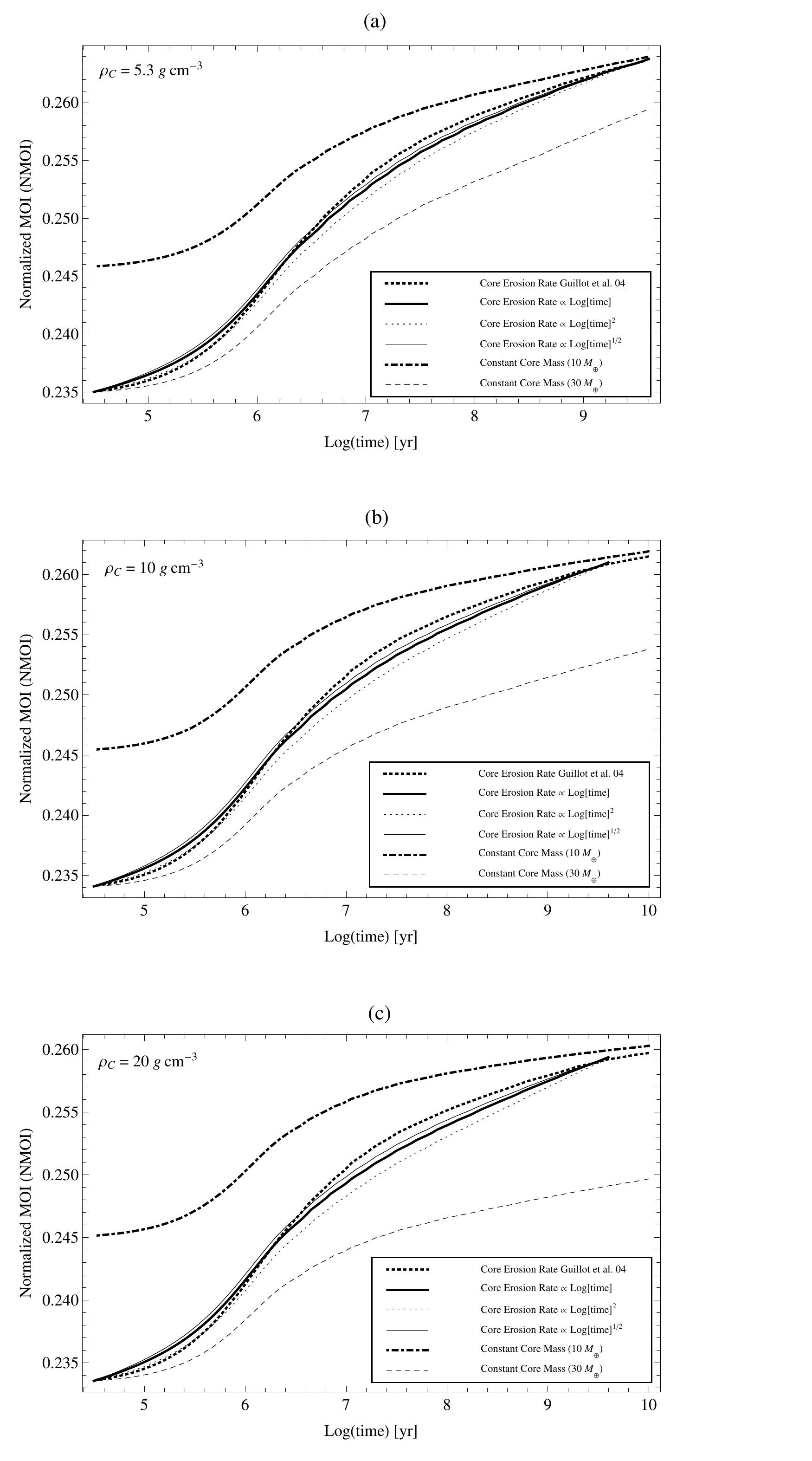} 
\caption{Results for Jupiter's NMOI as a function of time for the six different cases we consider. The identification of the different curves is the same as in figure 1. Results are shown for three different assumed core densities: 5.3 g cm$^{-3}$(a), 10 g cm$^{-3}$(b) and 20 g cm$^{-3}$(c) (see text for details). Jupiter's NMOI today is estimated to be between 0.2629 and 0.2645 (Helled et al., 2011). The NMOI derived by WC06 is about 0.236.
}
\end{center}
\end{figure}


\end{document}